\documentstyle[preprint,aps]{revtex}

\begin{document}

\draft

\preprint{gr-qc/9809030}
\tightenlines

\title{The quadratic spinor Lagrangian \\
is equivalent to the teleparallel theory}

\author{Roh Suan Tung
      \thanks {Electronic address : r.tung@lancaster.ac.uk}}

\address{
     Department of Physics, Lancaster University \\
     Lancaster LA1 4YB, United Kingdom }

\author{James M. Nester
       \thanks {Electronic address : nester@joule.phy.ncu.edu.tw}}

\address{
     Department of Physics and Center for Complex systems, National Central
University \\  Chungli 320, Taiwan, R.O.C.}

\date{23 March 1999}

\maketitle

\begin{abstract}
The quadratic spinor Lagrangian is shown to be equivalent to the
teleparallel/tetrad representation of Einstein's theory.  An important 
consequence is that the energy-momentum density obtained from this
quadratic spinor Lagrangian is essentially the same as the ``tensor''
proposed by M{\o}ller in 1961.
\end{abstract}


                                                             
\section*{Introduction} 

The Quadratic Spinor Lagrangian (QSL) formulation of General
Relativity (GR) \cite{NT95,TJ95,T96,Ro96,T99} has an appeal that goes  
beyond aesthetics.  One of its most promising features concerns
gravitational energy-momentum and its localization.

Identifying a suitable energy-momentum expression for the
gravitational field has long been an outstanding problem.  The usual
approaches lead to reference frame dependent pseudotensors; it seemed
that the best one could get was a quasilocal expression.  In contrast, 
in addition to links with the Witten type spinor formulation and an
associated positive energy proof, the QSL seems to yield a covariant
energy-momentum density \cite{NT95}.
This apparent covariance is here shown to be actually only cosmetic. 

In the earlier investigations the role of the spinor field used in   
this formulation had not been clarified.  In this present work it is 
shown that, at least for a large class of QSLs which we have         
considered, the spinor field is entirely an extra gauge field, which 
simply serves to give an attractive appearance to the formulas.      

Using a certain particularly simple QSL, (with the aid of a suitable 
gauge) we show that the whole formulation is equivalent to the
teleparallel (tetrad) reformulation of GR used by M{\o}ller in 1961  
\cite{Mo61}; consequently the associated energy-momentum density     
coincides with the energy-momentum ``tensor'' found by M{\o}ller.
This object {\em is} a tensor with regard to coordinate
transformations but {\em is not} a tensor with regard to local Lorentz
rotations of the frame.  Hence the associated energy-momentum
localization depends on a choice of Lorentz gauge.  Thus we conclude
that this QSL gives an energy-momentum density which actually depends
on a choice of Lorentz gauge.

Recall that
for Einstein's General Relativity there are various representations,
these include
(i) the metric, using coordinate frames,
(ii) orthonormal frames,
(iii) a teleparallel geometry, and
(iv) the quadratic spinor formulation.
Each representation reveals some insight and has some utility.  We
will briefly consider their relationships and compare the tensorial
nature of their associated energy-momentum expressions.  It turns out
that the last three representations are essentially equivalent.

The Hilbert Lagrangian density for GR is
${\cal L}_H=-\sqrt{-g} R$.
The traditional approach uses the metric coefficients in a coordinate
basis as the dynamic variables, so                                    
${\cal L}_H={\cal L}_H (g,\partial g, \partial\partial g)$.
Because of the second derivatives, this is not suitable for getting an
energy-momentum density.  However a certain (noncovariant) divergence
can be removed (without affecting the equations of motion
\cite{vacuum}) leading to Einstein's Lagrangian                       
${\cal L}_E={\cal L}_E(g,\partial g)={\cal L}_H-\hbox{div}$.
One can now apply the standard procedure and get the
canonical energy-momentum density.  It is known as the Einstein {\em
pseudotensor}; its value depends to a large extent on the coordinate
(``gauge'') choice.  No satisfying technique has been found to
separate the ``physics'' from the coordinate gauge.

\section*{M{\O}ller's tetrad/teleparallel representation}

An alternative is to use an
orthonormal frame (tetrad), a pioneer of this approach was          
M{\o}ller \cite{Mo61}. Let $g_{\mu\nu}=g_{ab}e^a{}_\mu e^b{}_\nu$,  
with $g_{ab}=\hbox{diag}(+1,-1,-1,-1)$,
and regard the Einstein-Hilbert Lagrangian as a function
${\cal L}_e(e,\partial e,\partial\partial e)$ of the tetrad
$e^a{}_\mu$. A suitable total divergence can again be removed yielding
 \begin{equation}
{\cal L}_M={\cal L}_M(e,\partial e)={\cal L}_e-\hbox{div},
\label{LM}
\end{equation}
a Lagrangian density which is first order in the derivatives of the
frame.   Now the standard canonical energy-momentum density
\begin{equation}
\sqrt{-g}T^\mu{}_\nu={\partial {\cal L}_M\over \partial \partial_\mu
e^a{}_\lambda}\partial _\nu e^a{}_\lambda-\delta^\mu_\nu{\cal L}_M,
\end{equation}
{\em is} a tensor (density)
under coordinate transformations, {\it but} it depends on the choice of
orthonormal frame (Lorentz gauge).  In other words it is not
``tensorial'' with respect to local ``rotations'' of the frame.
An alternate geometric viewpoint of this situation is to use a      
teleparallel formulation.                                           

Geometry includes, in general, the idea of parallel, which is          
determined by a connection.  {\it A priori} the connection could be    
independent of the metric.  Riemannian geometry (the standard type for 
GR) has a symmetric, metric compatible connection.  Parallel transport 
is then determined entirely by the metric.  One alternative is a       
teleparallel geometry (aka absolute parallel or Weitzenb\"ock          
geometry \cite{tele}) which has a connection with vanishing curvature. 
Parallel transport is then path independent.                           

The tetrad formulation of GR can be represented in terms of a          
teleparallel geometry.  This leads to the standard                     
Teleparallel Equivalent of GR, which has been referred to by           
several names including GRtele, GR$_{||}$ and TEGR.                    
(For further discussion of this theory and its applications see        
\cite{Mo61,Cho76,Me82,Ne89,Mie92,Ma94,Gro97,AP97,MGH98,NeYo99}         
and the references contained therein.) \                               
The idea is to introduce a new parallel
transport law.  This can be done via a simple construction:
(i) choose any orthonormal frame field,
(ii) define it to be parallel.
Then in this special OT (ortho-teleparallel) frame
the connection coefficients vanish:
 \begin{equation}
\Gamma^a{}_{b\mu}:=(\nabla_\mu e_b)^a=0;                               
\end{equation}
consequently the curvature vanishes in this and every other frame.
However the geometry is not trivial, for the components of the
torsion tensor,
\begin{equation}
T^a{}_{\mu\nu}=\partial_\mu e^a{}_\nu-\partial_\nu e^a{}_\mu
+\Gamma^a{}_{b\mu}e^b{}_\nu-\Gamma^a{}_{b\nu}e^b{}_\mu,
\end{equation}
in the OT frame reduce to
 $\partial_\mu e^a{}_\nu-\partial_\nu e^a{}_\mu$,
which is generally nonvanishing.  Being a tensor,  the torsion
will also be non-vanishing in any other reference frame.

Using this type of geometry, Einstein's GR theory can be obtained
from a Lagrangian quadratic in torsion:
 \begin{equation}
{\cal L}_T=\sqrt{-g}({1\over4}T^\alpha{}_{\mu\nu}T_\alpha{}^{\mu\nu}+
{1\over2}T^{\alpha\beta\mu}T_{\beta\alpha\mu}-
T^\alpha{}_{\alpha\mu}T^\beta{}_\beta{}^\mu).
\label{LT}
\end{equation}
The solutions to the field equations and the associated
energy-momentum tensor are now (teleparallel) gauge dependent.  Hence
the physics is represented by a whole gauge equivalence class of
teleparallel geometries \cite{sourcecoupling}.  This can be regarded  
just as a (sometimes quite useful) geometric reformulation of the     
usual tetrad formulation of GR.  The Lagrangian ${\cal L}_T$ is then
just an alternate ``more geometric'' interpretation of ${\cal L}_M$.
However, a strong case has been made for regarding this formulation as 
much more fundamental:  seeing it as a gauge theory for local          
translations, and in fact the ``correct'' way to understand GR as a    
gauge theory\cite{Cho76,Ma94,Gro97,AP97,MGH98}.                        

\section*{The quadratic spinor Lagrangian}

A few years ago (using some spinor-curvature identities \cite{NTZ94})
we found some quadratic spinor actions for GR \cite{NT95}.
 One of the simplest 
(recently we learned that a Lagrangian of the same form was used long
ago for anticommuting Majorana spinors in the context of supergravity
\cite{BMD77}) is dependent on a spinor valued one form $\Psi$:
\begin{equation}
 S[\Psi, \omega^{ab}]=
 \int{\cal L}_{\Psi} =
 \int 2 D\overline\Psi \gamma_5 D\Psi;
\end{equation}
here the covariant
differential,
$D\Psi:=d\Psi+\omega\Psi$,
includes the Clifford algebra
valued connection one-form
$\omega:=\textstyle{1\over4}\gamma_{ab}\omega^{ab}$.
(The Dirac matrix conventions are                              
$\gamma_{(a}\gamma_{b)}=g_{ab}$,                               
$\gamma_{ab}:=\gamma_{[a}\gamma_{b]}$,                         
$\gamma_5:=\gamma_0\gamma_1\gamma_2\gamma_3$.                  
We often omit the wedge $\wedge$;                              
for discussions of such ``clifform'' notation see              
\cite{Mi87,DMH91,Es91}.)                                       
This QSL satisfies
the spinor-curvature identity
\begin{equation}
{\cal L}_{\Psi}=2D\overline\Psi \gamma_5D\Psi
\equiv
2\overline\Psi\Omega\gamma_5\Psi+d[(D\overline\Psi)\gamma_5\Psi+
\overline\Psi\gamma_5D\Psi ],
\label{DPsi^2}
\end{equation}
where $\Omega={1\over4}\Omega^{ab}\gamma_{ab}=d\omega+\omega\omega$,
is the Clifford algebra valued curvature 2-form.
For the special case
$\Psi=\vartheta\psi$, which includes the orthonormal frame one-form
$\vartheta :=\gamma_a \vartheta^a=\gamma_a e^a{}_\mu dx^\mu$,
the rhs of (\ref{DPsi^2}) expands to
\begin{equation}
\overline\psi\psi \Omega^{ab}\wedge\eta_{ab}
 +\overline\psi\gamma_5 \psi \Omega_{ab}\wedge
\vartheta^a\wedge\vartheta^b
+d[D(\overline\psi\vartheta)\gamma_5 \vartheta\psi+
\overline\psi\vartheta \gamma_5 D(\vartheta\psi)],
\end{equation}
where we have introduced the convenient (Hodge) dual basis
$\eta^{a\dots}:=\ast (\vartheta^a\wedge\cdots)$.  For a
spinor field $\psi$, normalized according to
\begin{equation}
\overline\psi\psi=1,\qquad \overline\psi\gamma_5\psi=0, \label{norms}
\end{equation}
we get
\begin{equation}
{\cal L}_\psi=2D(\overline{\psi}\vartheta)\gamma_5 D(\vartheta\psi)
        \equiv \Omega^{ab}\wedge\eta_{ab}
        +  d[D(\overline{\psi}\vartheta)\gamma_5\vartheta\psi+
           \overline{\psi}\vartheta\gamma_5 D(\vartheta\psi)].
\label{R+dB}
\end{equation}
Since $\Omega^{ab}\wedge\eta_{ab}=-R\ast 1$,
this QSL differs from the standard Hilbert scalar curvature
Lagrangian  only by an exact differential.  In the action this
corresponds to a boundary term which does not affect the local
equations of motion \cite{spinorcoupling}.                         

\section*{Spinor gauge invariance of the QSL}

From the form of the Lagrangian (\ref{R+dB}), the QSL action for an
extended region actually depends on the (normalized) spinor field only
through the boundary term, not locally.  A change of the spinor field 
within the interior of the region will leave the action unchanged.    
Consequently the Dirac spinor field $\psi$ has complete local gauge
invariance subject to the two restrictions (\ref{norms}).  This 6 real
parameter spinor gauge freedom can be represented in the form
$\psi=U\psi_0$ where $\psi_0$ is a normalized Dirac spinor with
constant components and $U$ is the Dirac spinor representation of a
Lorentz transformation.  Thus the gauge freedom of the normalized
spinor field is a kind of local Lorentz gauge freedom.  Considering
the scalar curvature term in the Lagrangian (\ref{R+dB}), it can be
recognized that the theory also has the usual local Lorentz gauge
freedom associated with transformations of the orthonormal frame.
Hence there appears to be two Lorentz gauge freedoms here.          
But are they really independent?

Considering the covariant appearance of the boundary term in
(\ref{R+dB}) this seems doubtful.  Usually we regard a transformation
of Lorentz frame as inducing associated transformations on the
components of all tensors and spinors.  Under such a transformation
the Lagrangian boundary term is a Lorentz invariant.  But now we are
contemplating independent transformations of the spinor and frame
field.  How does the boundary term behave?

The boundary term is
 \begin{equation}
(D\overline{\psi}\gamma_a\gamma_5\gamma_b\psi-
           \overline{\psi}\gamma_a\gamma_5\gamma_b D\psi)
\vartheta^a\wedge\vartheta^b
+(\overline{\psi}\gamma_a\gamma_5\gamma_b\psi-
           \overline{\psi}\gamma_b\gamma_5\gamma_a \psi)
D\vartheta^a\wedge\vartheta^b .
\end{equation}
Let us consider a gauge transformed spinor field
$\psi'=U\psi$. Then $\overline{\psi'}=\overline{\psi}U^{-1}$,
$D\psi'=UD\psi$ and $D\overline{\psi'}=D(\overline{\psi})U^{-1}$.
The gauge transformed boundary term then becomes
\begin{eqnarray}
(D\overline{\psi}U^{-1}\gamma_a U\gamma_5 U^{-1}\gamma_b U\psi-
  \overline{\psi}U^{-1}\gamma_a U\gamma_5 U^{-1}\gamma_b UD\psi)
&&\vartheta^a\wedge\vartheta^b \nonumber\\
+(\overline{\psi}U^{-1}\gamma_a U\gamma_5 U^{-1}\gamma_b U\psi-
  \overline{\psi}U^{-1}\gamma_b U\gamma_5 U^{-1}\gamma_a U\psi)
&&D\vartheta^a\wedge\vartheta^b.
\end{eqnarray}
The unitary transformations on the gammas induce Lorentz
transformations, $U^{-1}\gamma_a U=\gamma_c L^c{}_a$, on the
orthonormal frame indices.  Such a transformation is entirely
equivalent to applying the
transformation $\vartheta'{}^c=L^c{}_a\vartheta^a$ to the orthonormal
frame alone.  Hence the boundary term really has one
physically independent Lorentz gauge freedom.

\section*{Equivalence of the Lagrangians}              

Without losing any physics, we can confine our attention to
representations where the spinor field and the orthonormal frame are
tied together.  A convenient choice fixing one of the Lorentz gauges
is $d\psi=0$; in other words the components of $\psi$ are constant in 
the present frame.  This locks the spinor and the orthonormal frame
together.  The pair then still retain the other Lorentz gauge
freedom.  For a general analysis there may be some advantage in
regarding this condition as binding the orthonormal frame to the
spinor field which (viewed as a geometric object, not as a set of
components) retains its Lorentz gauge freedom.  However for our immediate
needs it is more perspicuous to consider the condition as tying        
the spinor to the orthonormal frame, with the latter still retaining
its own local Lorentz gauge freedom.

We could establish the equivalence by directly
expanding the Lagrangian
\begin{equation}
{\cal L}_\psi =2D(\overline{\psi}\vartheta)\gamma_5 D(\vartheta\psi)
=2(d(\overline{\psi}\vartheta)+\overline{\psi}\vartheta\omega)\gamma_5
   (d(\vartheta\psi)+\omega\vartheta\psi),
\label{Lpsi}
\end{equation}
using $d\psi=0$,                                       
however an indirect calculation is more efficient.
We consider the boundary term on the rhs of the
Lagrangian (\ref{R+dB}).
With the gauge choice
$d\psi=0$ in the present frame, we find
 \begin{eqnarray}
[d(\overline{\psi}\vartheta)+\overline{\psi}\vartheta\omega]
\gamma_5\vartheta\psi &+&
  \overline{\psi}\vartheta\gamma_5[ d(\vartheta\psi)+\omega\vartheta\psi]
\equiv
\overline\psi (d\vartheta+\vartheta\omega)\gamma_5 \vartheta\psi
+\overline\psi\vartheta\gamma_5 (d\vartheta+\omega\vartheta)\psi\nonumber\\
&\equiv&-\overline\psi\psi \omega^{ab}\wedge \eta_{ab}
+\overline\psi\gamma_5 \psi\omega_{ab}\wedge \vartheta^a\wedge\vartheta^b
+\overline\psi\gamma_5 \psi d\vartheta^a\wedge\vartheta_a.
\end{eqnarray}
Hence, taking into account the spinor field normalization conditions
(\ref{norms}), the Lagrangian (\ref{R+dB}) is equivalent to
\begin{equation}
{\cal L}\equiv(d\omega^{ab}+\omega^a{}_c\wedge\omega^{cb})\wedge\eta_{ab}
-d(\omega^{ab}\wedge\eta_{ab})\equiv
\omega^{a}{}_c\wedge\omega^{cb}\wedge\eta_{ab}
+\omega^{ab}\wedge d\vartheta^c\wedge\eta_{abc}.
\label{ww+wdo}
\end{equation}
If we vary the connection independently, we find (with no source)
a relation (equivalent to vanishing torsion) which shows that the connection
is a particular linear combination of $d\vartheta$.  This relation is just
the usual expression for the orthonormal frame (Riemannian) connection
coefficients $\omega_{abc}=\textstyle{1\over2}(C_{cab}-C_{bca}-C_{abc})$,
where $C^a{}_{bc}:=-d\vartheta^a(e_b,e_c)$.  Inserting these values into the
rhs of (\ref{ww+wdo}) gives an explicit form of the M{\o}ller
Lagrangian (\ref{LM}) for tetrad gravity:
\begin{equation}
{\cal L}_M=\left({1\over4}C_a{}^{bc}C^a{}_{bc}-C_c{}^{cb}C^a{}_{ab}
+{1\over2}C_{bac}C^{abc}\right) *1,
\end{equation}
The equivalent covariant description in terms of teleparallel
geometry (\ref{LT}) readily follows since the torsion 2-form
\begin{equation}
{1\over2} T^a{}_{\mu\nu}dx^\mu\wedge dx^\nu=T^a
:=D\vartheta^a:=d\vartheta^a+\omega^a{}_b\wedge\vartheta^b,
\end{equation}
reduces, in an OT frame, to $d\vartheta^a$.
Hence we have established
an equivalence between the QSL (\ref{Lpsi}) and the
tetrad/teleparallel representations of GR.  As we have mentioned, 
these latter representations were used by M{\o}ller to construct a
gravitational energy-momentum density.
Consequently the corresponding energy-momentum localization
for the QSL and the M{\o}ller representations
are also equivalent.

\def\N{\not\!\! N}

\section*{Energy-momentum quasilocalization}

It is instructive to consider this point from the QSL expressions.
The energy-momentum density can be identified with the Hamiltonian.
The Hamiltonian can be constructed from the action by choosing a
timelike evolution vector field $N$ such that $i_N dt = 1$ and
splitting the action:
$S=\int {\cal L} =  \int dt \int i_N{\cal L}$.
Applied to ${\cal L}_\psi$ this procedure leads \cite{NT95} to the
 4-covariant QSL Hamiltonian 3-form \cite{witham} (i.e., the        
Noether translation generator along $N$)
\begin{equation}
{\cal H}(N)
=2[ D(\overline\psi\N)\gamma_5 D(\vartheta\psi)+
 D(\overline\psi\vartheta)\gamma_5 D(\N\psi)].
\label{ham}
\end{equation}
A notable feature of this QSL Hamiltonian
 is that it is already asymptotically $O(1/r^4)$;                   
consequently, its integral will be finite and  its variation        
will have an $O(1/r^3)$ boundary term which will vanish asymptotically, 
so there is no need for any further boundary term adjustment.       
In fact the                                                         
Hamiltonian expression (\ref{ham}) could have been obtained from
the usual
(linear in the Einstein tensor) Hamiltonian by adding a certain total
differential (although important for the value of energy-momentum such
a total differential does not effect the equations of motion), as the
following identity reveals:
\begin{equation} {\cal H}(N)\equiv
 2\overline\psi\psi N^\mu G_{\mu\nu}*\!\vartheta^{\nu}+
 2d[\overline\psi\N\gamma_5 D(\vartheta\psi)+ D(\overline\psi\vartheta)
\gamma_5 \N\psi].
 \end{equation}
This identity also shows that the
derivatives of $\psi$ itself are not so important --- up to an exact
differential ${\cal H}(N)$ is algebraic in $\psi$ --- rather that
these factors arrange for the correct quadratic connection terms.

The Hamiltonian (\ref{ham}) looks covariant but what does it mean
physically?
When the constraint equations are satisfied, the value of the
Hamiltonian is given by the boundary term.   In the $d\psi=0$ gauge
the Hamiltonian boundary term is
\begin{eqnarray}
\overline\psi\N\gamma_5 D(\vartheta\psi)+
 D(\overline\psi\vartheta)\gamma_5 \N\psi
&=& N^c \overline\psi[\gamma_c\gamma_5(d\vartheta+\omega\vartheta)
+(d\vartheta+\vartheta\omega)\gamma_5\gamma_c]\psi \nonumber\\
&=&-N_c\overline\psi\gamma_5\psi d\vartheta^c
-N^c\overline\psi\psi\epsilon_{abcd}\omega^{ab}\wedge\vartheta^d.
\label{bound}
\end{eqnarray}
Hence, with the normalizations (\ref{norms}), we find that the value
of the boundary term is
\begin{equation}
\omega^{ab}\wedge i_N\eta_{ab}=N^c\omega^{ab}\wedge\eta_{abc}
=N^c\Gamma^{ab}{}_d\delta^{efd}_{abc}\eta_{ef},
 \end{equation}
a well known expression for the superpotential \cite{CN99} associated 
with M{\o}ller's energy-momentum ``tensor''.  This superpotential is a
tensor with respect to  coordinate transformations but is not
tensorial with respect to the
local Lorentz gauge freedom.  Without some gauge fixing condition the
M{\o}ller energy-momentum ``tensor'', and likewise the QSL formulation
{\em does not} determine a well defined gravitational energy
localization.  M{\o}ller realized the need for some Lorentz gauge
condition and even proposed one \cite{Mo61} (which did not prove to be
so satisfactory).  More recently a gauge condition has been proposed
which would determine certain {\em special orthonormal frames}
\cite{Ne92}.  However, both of these conditions depend (like typical
gauge conditions) on the solution
of a partial differential equation; hence the gauge fixed
Lorentz frames are inherently nonlocal.  Using such gauge fixed frames
will not yield a true
{\em local} energy momentum density.

\section*{Discussion}

In summary, we have demonstrated that a particular quadratic
spinor Lagrangian (QSL) is equivalent
to the tetrad/teleparallel version of Einstein's GR theory.
The corresponding energy-momentum density is equivalent to
M{\o}ller's 1961 ``tensor''.

Our analysis raises a question.                                     
First, to what extent is our conclusion dependent on the particular
choice of QSL (\ref{Lpsi})?
 Consider any Lagrangian, quadratic in the
derivatives of some field $W$ (spinor or otherwise), which
differs from the Hilbert Lagrangian only by an exact differential:
\begin{equation}
{\cal L}_W=(DW)^2\equiv -R\ast 1+d(\dots).
\end{equation}
Then $W$ is necessarily a pure local gauge field --- since the action
depends on $W$ {\em only} through a boundary term.  In practice we only
succeeded in getting the scalar curvature in this type of identity by using
spinor fields.  All the spinor curvature identities \cite{NTZ94}
involving the scalar curvature (the Hilbert Lagrangian) which we
found, differed from (\ref{Lpsi}) only by torsion terms, which would
make no contribution to the field equations in the sourceless case.   
(For the case with sources which couple to the Riemannian connection, 
we can recover Einstein's theory by suppressing the torsion using a   
Lagrange multiplier term.) \                                          
Hence all QSLs leading to Einstein's
theory are essentially equivalent to the particular one discussed
here.

It is well known that the M{\o}ller energy-momentum ``tensor'' and the
associated superpotential depends on the local Lorentz frame gauge.
The total energy-momentum within a finite region actually depends only
on the integral of the superpotential over the boundary --- hence it
depends on the local frame only through the values on the boundary.
Similarly, the spinor field is purely a local gauge field; only its  
value on the boundary influences the calculation of energy-momentum
within a region.  From either representation we once again see
energy-momentum as quasilocal \cite{CNT95,CN99}, depending on the fields 
and the gauge choice on the boundary.

We have shown that the role of the spinor field in the QSL
representation is essentially cosmetic; it allows a neat             
alternate version of the tetrad/teleparallel representation of GR.
(Actually the representations are not quite equivalent because of
the 2 to 1 relation between the spinor and frame gauge groups.)
This is not to say that the QSL representation is useless.  The QSLs
can essentially replace an orthonormal frame with a spinor field     
(this works because a normalized spinor field determines an          
orthonormal frame up to an overall constant Lorentz transformation)  
which may have some advantages.  One is that, like any other new     
representation, the QSL suggests certain                             
generalizations (e.g., to complex self dual representations)
\cite{G}. 
This
present work does not preclude the possibility that a QSL would lead
to a representation with genuinely new features.  That could happen if
it was not directly connected to the Hilbert scalar curvature
Lagrangian.

\section*{Acknowledgments}                                          
We would like to thank the referees for their helpful suggestions.  
This work was supported by the National Science Council             
of the R.O.C. under grant No. NSC87-2112-M-008-007,
NSC88-2112-M-008-018.


\begin{references}
\bibitem{vacuum} For the body of our work we consider the source free 
`vacuum' theory.  The issue of coupling to sources is noted later.
\bibitem{sourcecoupling}
Concerning the inclusion of sources in the field equations,           
the proper source coupling is such that a test particle without spin  
will only feel the Riemannian structure.  In GR sources with spin     
couple to the connection;      
their standard GR coupling is not          
equivalent to `minimal coupling' with respect to the teleparallel     
geometry.
\bibitem{spinorcoupling} For the equation of motion with sources,  
the equivalent to the GR source coupling                              
is reproduced by incorporating the metric (via the tetrad) into the   
source Lagrangian.  In some cases that may be easily translated into  
an equivalent coupling to the spinor field $\psi$.  However, we do not 
expect to find a general prescription for coupling the QSL spinor
field to a source in a way that would be compatible with the spinor's
local Lorentz gauge freedom.
\bibitem{witham} This Hamiltonian is similar to                       
but distinct from the one associated                                  
with the Witten positive energy proof\cite{NT95,CNT95}.
\bibitem{G}   
Some recent works \cite{MM99,Macias96} have
noted a connection between the QSL formulation 
and chiral supergravity.
For a connection with Ashtekar's new variables see 
 \cite{Wallner95}.
\end{references}
\end{document}